\documentclass[prl,showpacs,twocolumn]{revtex4-1}
\usepackage{graphicx}
\usepackage{color}

\begin{document}

\title{Quantum Thermal Rectification to design thermal diodes and transistors}
\author{Karl Joulain, Youn\`es Ezzahri and Jose Ordonez-Miranda}
\email{Corresponding author : karl.joulain@univ-poitiers.fr}
\affiliation{Insitut Pprime, CNRS, Universit\'e de Poitiers, ISAE-ENSMA, F-86962 Futuroscope Chasseneuil, France}

\date{\today}
\begin{abstract}
We study in this article how heat can be exchanged between two level systems (TLS) each of them being coupled to a thermal reservoir. Calculation are performed solving a master equation for the density matrix using the Born markov-approximation. We analyse the conditions for which a thermal diode and a thermal transistor can be obtained as well as their optimization.
\end{abstract}

\maketitle

\section{Introduction}

Global warming and limited energy issues have increased the interest in the energy management and in particular heat losses. Indeed, heat wasted in energy production processes and thermal machines could in principle be better used in many applications if it could be guided or transport in a similar way as electricity. However, if heat pipes have been proved to be good candidates for thermal guiding, there exists few devices at the moment that can switch or amplify heat as is the case for electricity.

In electricity, the development of diodes \cite{lashkaryov_investigations_1941} and transistors \cite{bardeen_transistor_1998} have led to its control at the scale of the electron, leading to the emergence of electronics. One can therefore wonder whether heat could be managed in the same way, if the thermal equivalent of these two objects would exist. In the last decade, several works have focused on the development of thermal rectifiers, i.e devices for which the thermal fluxes flowing through them is different in magnitude when the temperatures are inverted at their ends. Thus, phononic \cite{terraneo_controlling_2002,li_thermal_2004,li_interface_2005,chang_solid-state_2006,hu_asymmetric_2006,yang_thermal_2007,hu_thermal_2009,pereira_sufficient_2011,zhang_thermal_2011,roberts_review_2011,garcia-garcia_thermal_2014}  and electronic \cite{roberts_review_2011,segal_single_2008} thermal diodes or rectifiers have been developed, that have later led to the proposition of thermal transistors \cite{wang_thermal_2007,chung_lo_thermal_2008}. Later, these concepts have been extended to the case of thermal radiation both in the near field \cite{otey_thermal_2010-1,basu_near-field_2011,ben-abdallah_phase-change_2013} and far field \cite{van_zwol_emissivity_2012,ito_experimental_2014,nefzaoui_simple_2014,nefzaoui_radiative_2014,joulain_radiative_2015}. The most interesting results have been found through the use of phase change thermochrome materials \cite{huang_thermal_2013}, such as VO$_2$\cite{morin_oxides_1959,rini_photoinduced_2005}. Recently, thermal transistors have been designed using similar properties \cite{ben-abdallah_near-field_2014,joulain_modulation_2015}.

In the last years, individual quantum systems, such as classical atoms \cite{brune_quantum_1996,maunz_cavity_2004} or artificial ones, as is the case of quantum dots \cite{claudon_-chip_2009,dousse_ultrabright_2010}, have been proposed to develop photon rectifiers \cite{yu_complete_2009,mascarenhas_quantum_2014,mascarenhas_quantum_2015}, transistors \cite{hwang_single-molecule_2009,astafiev_ultimate_2010} or even electrically controlled phonon transistors \cite{jiang_phonon_2015}. Moreover, as quantum systems are always coupled to the environment, the question of how heat is transferred through a set of quantum systems in interaction naturally has arisen \cite{manzano_quantum_2012,bermudez_controlling_2013,pumulo_non-equilibrium_2011}  and led to several works on thermal rectification \cite{scheibner_quantum_2008,pereira_symmetry_2009,werlang_optimal_2014,chen_thermal_2015}.

The goal of this article is to use elementary quantum objects, such as two-level systems (TLS) related to thermal baths, for developing thermal diodes and thermal transistors. To do so, we will use the classical quantum thermodynamics formalism proposed by Lindblad  that is based on the resolution of a master equation for the density matrix. We show, following the work of Werlang et al. \cite{werlang_optimal_2014}, that 2 TLS can easily make a thermal diode and that 3 TLS can make a thermal transistor. These three TLS related to thermal reservoirs are equivalent to the three entries of a bipolar electronic transistor. It is shown that a thermal current imposed at the base can drive the currents at the two other entries of the system. 

\section{Theory}
We consider in the following, that TLS are connected to a thermal bath and that can be coupled one to each other. Two configurations are studied in this article:  2 TLS coupled to each other make a thermal diode, whereas 3 coupled TLS make a thermal transistor.
\subsection{Thermal diode}
\begin{figure}
\begin{center}
\includegraphics[width=7cm]{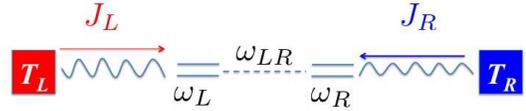}
\caption{Quantum thermal diode made up of 2 TLS coupled with each other and connected to a thermal bath.}
\label{systemdiode}
\end{center}
\end{figure}

The system under consideration consists of two coupled TLS, each of them related to a thermal bath, as depicted in Fig.\ref{systemdiode}.

The two TLS are labeled with the letters $L$ (left) and $R$ (right), which is also the case of the temperature of the thermal baths related to TLS.  We use the strong-coupling formalism developed by Werlang {\it et al.} \cite{werlang_optimal_2014}. Each of the TLS is caracterized by an angular frequency $\omega_L$ or $\omega_R$. The coupling between the two TLS has the typical angular frequency $\omega_{LR}$. The hamiltionian of the system is (in $\hbar=1$ units) 
\begin{equation}
\label{ }
H_S=\frac{\omega_L}{2}\sigma_z^L+\frac{\omega_R}{2}\sigma_z^R+\frac{\omega_{LR}}{2}\sigma_z^L\sigma_z^{R},
\end{equation}
where $\sigma_z^P$ ($P=L,R$) is the Pauli matrix $z$, whose eigenstates for the system $P$ are the states $\uparrow$ and $\downarrow$. $H_S$ eigenstates are given by the tensorial product of the individual TLS states, so that we have 4 eigenstates labeled as $\left|1\right>=\left|\uparrow\uparrow\right>$, $\left|2\right>=\left|\uparrow\downarrow\right>$, $\left|3\right>=\left|\downarrow\uparrow\right>$, $\left|4\right>=\left|\downarrow\downarrow\right>$. The coupling between the TLS and the thermal bath $P$ constituted of harmonic oscillators \cite{caldeira_quantum_1983}, is based on the spin-boson model in the $x$ component
$H_{\rm TLS-bath}^P=\sigma_x^P\sum_k g_k(a_k^P a_k^{P\dag})$.
The two reservoirs $P$ have their Hamiltonians equal to
$H_{\rm bath}^P=\sum_k\omega_ka_k^{P\dag}a_k^P$.
This modeling implies that baths can only flip one spin at a time. There are therefore 4 authorized transitions. Transitions $1\leftrightarrow3$ and $2\leftrightarrow4$ are induced by the thermal bath $L$, whereas transitions $1\leftrightarrow2$ and $3\leftrightarrow4$ are induced by the thermal bath $R$. Transitions $1\leftrightarrow4$ and $2\leftrightarrow3$ are forbidden.

The system state is described by a density matrix $\rho$, which obeys a master equation. In the Born-Markov approximation, it reads
\begin{equation}
\label{master}
\frac{d\rho}{dt}=-i[H_s,\rho]+{\cal{L}}_L[\rho]+{\cal{L}}_R[\rho].
\end{equation}
As in \cite{werlang_optimal_2014,breuer_theory_2002}, the Lindbladians ${\cal{L}}_P[\rho]$ are written for an Ohmic bath according to classical textbooks \cite{leggett_dynamics_1987,breuer_theory_2002}, so that we take the expression
\begin{eqnarray}
{\cal{L}}_P[\rho]&=&\sum_{\omega>0}{\cal{I}}(\omega)(1+n_\omega^P)\nonumber\\
&\times&\left[A_P(\omega)\rho A^+_P(\omega)-\frac{1}{2}\left\{\rho,A^+_P(\omega)A_P(\omega)\right\}\right]\\
&+&{\cal{I}}(\omega)n_\omega^P\left[A^+_P(\omega)\rho A_P(\omega)-\frac{1}{2}\left\{\rho,A_P(\omega)A^+_P(\omega)\right\}\right]\nonumber
\end{eqnarray}
of \cite{werlang_optimal_2014}, where 
\begin{equation}
\label{ }
n_\omega^P=\frac{1}{e^{\hbar\omega/k_bT-1}},
\end{equation}
and 
\begin{equation}
\label{ }
A_P(\omega)=\sum_{\omega=\epsilon_j-\epsilon_i}\left|i\right>\left<i\right|\sigma_x^P\left|j\right>\left<j\right|.
\end{equation}
We now consider a steady state situation. We define 
\begin{equation}
\label{currentdef}
{\rm Tr}(\rho {\cal{L}}_P[\rho])=J_P,
\end{equation}
the heat current injected by the bath $P$ into the system.  Averaging the master equation, we find $J_L+J_R=0$, in accordance with the energy conservation.

\begin{figure}
\begin{center}
\includegraphics[width=8cm]{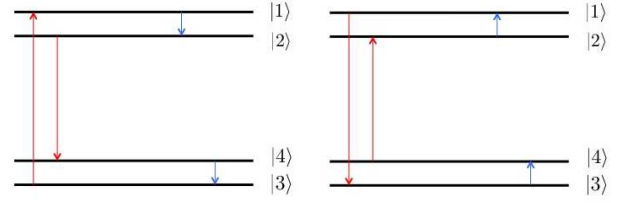}
\caption{Energy levels and transitions in the case $\omega_L=1$, $\omega_R=0$ and $\omega_{LR}=0.1$. The arrow directions shows the balance of the authorized transition between levels. Left : $T_L>T_R$. Right $T_L<T_R$. }
\label{level_diode}
\end{center}
\end{figure}

\begin{figure}
\begin{center}
\includegraphics[width=8cm]{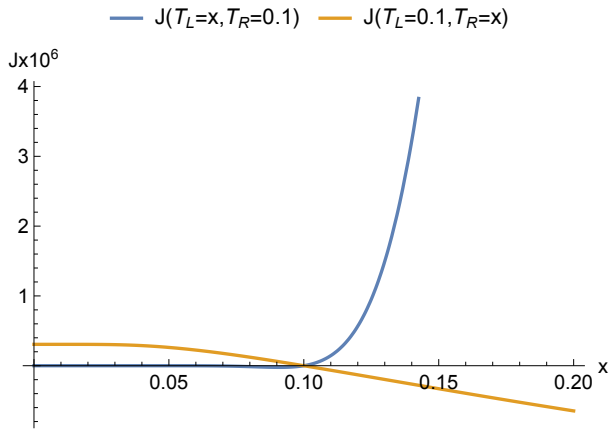}
\caption{$J_L(T_L,T_R)$ in the case $\omega_L=1$, $\omega_R=0$ and $\omega_{LR}=0.1$ with $T_R=1$}
\label{Flux1001}
\end{center}
\end{figure}

The master equation is a system of four equations on the diagonal elements $\rho_{ii}$. If we introduce the net decaying rate from state $\left|i\right>$ to the state $\left|j\right>$, due to the coupling with bath $P$, with the help of Bose-Einstein distribution $n_\omega^P=(e^{\omega/T_P}-1)^{-1}$ (in $k_b=1$ units):
$\Gamma_{ij}^P=\omega_{ij}\left[\left(1+n_\omega^P\right)\rho_{ii}-n_\omega^P\rho_{jj}\right]=-\Gamma_{ji}^P$,
the master equation yields
\begin{eqnarray}
\dot{\rho}_{11} & = & 0 =  \Gamma_{31}^L + \Gamma_{21}^R, \nonumber  \\
\dot{\rho}_{22} & = & 0 = \Gamma_{42}^L + \Gamma_{12}^R,\nonumber \\
\dot{\rho}_{33} & = & 0 = \Gamma_{13}^L  + \Gamma_{43}^R, \label{rhodiode}\\
\dot{\rho}_{44} & = & 0 = \Gamma_{24}^L + \Gamma_{34}^R, \nonumber
\end{eqnarray}
from which it can be deduced that 
\begin{equation}
\label{ }
\Gamma_{31}^L=\Gamma_{24}^L=\Gamma_{12}^R=\Gamma_{43}^R=\Gamma.
\end{equation}
The definition of the thermal currents (\ref{currentdef}) gives then the final expression of the thermal currents
\begin{equation}
\label{ }
J_L=-J_R=2\omega_{LR}\Gamma
\end{equation}
As an example, let us consider the example of a system where $\omega_L=1$, $\omega_R=0$ and $\omega_{LR}=0.1$. The energy levels and the authorized transitions are depicted in Fig.\ref{level_diode}
When $T_L>T_R$, the left reservoir populates level $\left|1\right>$ from level $\left|3\right>$ through transition $1\leftrightarrow 3$. Level $\left|1\right>$ de-excitates through level $\left|2\right>$ by transfering energy to reservoir $R$. Level $\left|2\right>$ de-excitates through level $\left|4\right>$ by transfering energy to reservoir $L$ and finally level $\left|4\right>$ de-excitates through level $\left|3\right>$ by transfering energy to reservoir $R$. If $T_L<T_R$, the energy transfers are reversed. Now imagine that $T_L$ is of the order of the transition energies, whereas $T_R$ is much lower. Then, energy will easily flow from reservoir $L$ to reservoir $R$ according to the process described above. On the contrary, if $T_R$ is much lower than the transition energies and $T_L<T_R$ then the energy transfer is poor since excitation by reservoir $R$ through transition $4\leftrightarrow3$ and $2\leftrightarrow1$ is low. Hence, if we study the flux $J_L(T_L,T_R)$ with $T_R$ fixed at a value lower than the transition energies (for example $T_R=0.1$, Fig. \ref{Flux1001}), we see that the flux is close to 0 when $T_L<T_R$ . 
When $T_L$ is increased to values larger than $T_R$, the current inceases until saturation at high temperatures. The calculation of $\Gamma$, which gives the current, can be achieved by solving the system of equations on the populations ($\ref{rhodiode}$). Note that the system of equations are not totally independent since the fourth equation is actually the sum of the three others. One has to use the fact that the trace of the density matrix is equal to 1 (Tr[$\rho$]=1). The exact expression of $\Gamma$ can be found in \cite{werlang_optimal_2014}. In the case studied here, this expression can be simplified and the current reads
\begin{equation}
\label{ }
J_L\approx \frac{\omega_L\omega_{LR}}{2}\frac{e^{-\omega_{LR}/T_L}}{\cosh(\omega_L/T_L)}
\end{equation}
where the transition from low current values, at low $T_L$, to high current values, at higher $T_L$, can be seen. 

Let us note that the system proposed here constitutes a passive thermal switch at low temperature. As long as $T_L$ is larger than $T_R$, the current in the structure is important and the thermal contact is good between the reservoirs $L$ and $R$. However, when the temperature $T_L$ reduces to values below $T_R$, the thermal current is drastically lowered, so that it can be seen as switched off. This system could therefore be used to isolate objects from a cold environment while it would be thermally linked to a hot environment. In a case of an environment with temperatures oscillating between high and low values, this simple quantum system can be seen as a passive heater and a thermal rectifier, i.e that heat flow through it depends on the direction of the heat flux. 

\begin{figure}
\begin{center}
\includegraphics[width=8cm]{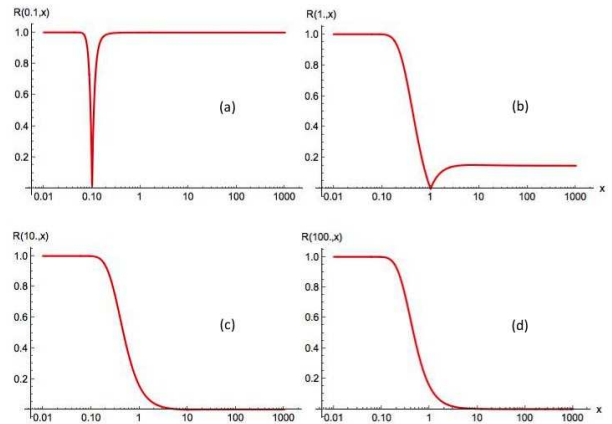}
\caption{Rectification ratio $R(T_L,T_R)$ variations with $T_L$ for different values of $T_R$. (a) $T_R=0.1$. (b) $T_R=1$. (c) $T_R=10$. (d) $T_R=100$.  }
\label{rectif}
\end{center}
\end{figure}

There is actually another way to quantify the rectification of a system. This is the ratio between the sum of the fluxes through the system when the temperatures are reversed and the maximum of these 2 fluxes
\begin{equation}
\label{Rectification}
R(T_L,T_R)=\frac{|J_L(T_L,T_R)+J_L(T_R,T_L)|}{Max(|J_L(T_L,T_R)|,|J_L(T_R,T_L)|)}
\end{equation}
The rectification ratio $R(T_L,T_R)$ variations with $T_L$ for different $T_R$ are represented in Fig. \ref{rectif}.
When $T_R$ is small enough ($T_R<1$), rectification is strong except for values of $T_L$ very close to those of $T_R$. When $T_R$ is larger, rectification is smaller, even for $T_L$ values that are greatly different from $T_R$. We note in particular, that rectification is low for high $T_R$ temperature. In this later case, there is no rectification, because heat transfer can occur with both reservoir  with  help of the energy transitions presented above. However, when $T_R$ is fixed, and $T_L$ goes to 0, then $J_L(T_R,T_L)$ tends to 0. As well, rectification rises to 1. This kind of device can thus be seen as a thermal diode, since the heat current through the system is nonzero when the heat flux is in a given direction and 0 when it is in the opposite one.

This type of system paves the way to develop more complicated ones. For example, it is well known that electronic transistors as the bipolar ones, can be made up NPN et PNP junctions whereas it well known that the PN junction constitutes a diode. One can therefore wonder if it is also possible to conceive a transistor with the elementary quantum system that constitutes the thermal diode that we have just studied in this section. This is the subject of the next section.

\subsection{Thermal transistor}

\begin{figure}
\begin{center}
\includegraphics[width=7cm]{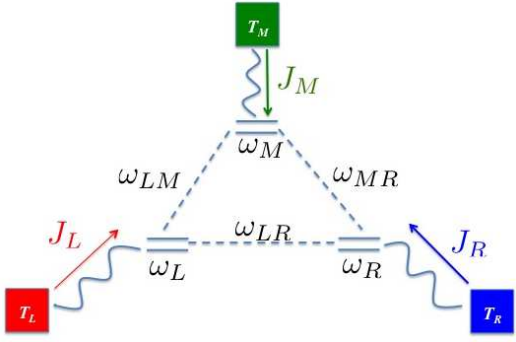}
\caption{Quantum system made up of 3 TLS coupled with each other and connected to a thermal bath.}
\label{system}
\end{center}
\end{figure}

The system studied in this part is constituted of three TLS coupled with each other, each of them being connected to a thermal bath (Fig. \ref{system}). This system is therefore similar to the previous one with one supplementary TLS and reservoir.
The three TLS are now labeled with the letters $L$ (left), $M$ (medium), and $R$ (right), as well as the temperature of the thermal baths involved. As in the previous part, we use the strong-coupling formalism developed by Werlang {\it et al.} \cite{werlang_optimal_2014}. 
Similarly, TLS can be in the up state $\uparrow$ or in the down one $\downarrow$. The Hamiltonian of the system is (in $\hbar=1$ units) 
\begin{equation}
\label{ }
H_S=\sum_{P=L,M,R}\frac{\omega_P}{2}\sigma_z^P+\sum_{P,Q=L,M,R \ P\neq Q}\frac{\omega_{PQ}}{2}\sigma_z^P\sigma_z^{Q}
\end{equation}
$\omega_P$ denotes the energy difference between the two spin states, whereas $\omega_{PQ}$ stands for the interaction between the spin $P$ and the spin $Q$. Following the preceding part on the quantum thermal diode, we have eight eigenstates labeled as $\left|1\right>=\left|\uparrow\uparrow\uparrow\right>$, $\left|2\right>=\left|\uparrow\uparrow\downarrow\right>$, $\left|3\right>=\left|\uparrow\downarrow\uparrow\right>$, $\left|4\right>=\left|\uparrow\downarrow\downarrow\right>$, $\left|5\right>=\left|\downarrow\uparrow\uparrow\right>$, $\left|6\right>=\left|\downarrow\uparrow\downarrow\right>$, $\left|7\right>=\left|\downarrow\downarrow\uparrow\right>$ and $\left|8\right>=\left|\downarrow\downarrow\downarrow\right>$. There are now 12 authorized transitions. The left bath ($L$) induces  the transitions $1\leftrightarrow5$, $2\leftrightarrow 6$, $3\leftrightarrow7$, and $4\leftrightarrow8$, the middle one ($M$) drives the transitions $1\leftrightarrow3$, $2\leftrightarrow 4$, $5\leftrightarrow7$, and $6\leftrightarrow8$. The right bath ($R$) triggers the transitions $1\leftrightarrow2$, $3\leftrightarrow 4$, $5\leftrightarrow6$, and $7\leftrightarrow8$. All other transitions flipping more than one spin are forbidden. 

The master equation fulfilling the density matrix, in the Born-Markov approximation, reads
\begin{equation}
\label{master}
\frac{d\rho}{dt}=-i[H_s,\rho]+{\cal{L}}_L[\rho]+{\cal{L}}_M[\rho]+{\cal{L}}_R[\rho].
\end{equation}
We now go to the steady state situation. Averaging the master equation, we find $J_L+J_M+J_R=0$, in accordance with the energy conservation.

The master equation is a system of eight equations on the diagonal elements $\rho_{ii}$. Introducing the net decaying rate from state $\left|i\right>$ to the state $\left|j\right>$ due to the coupling with bath $P$,
the master equation becomes
\begin{eqnarray}
\dot{\rho}_{11} & = & 0 = \Gamma_{51}^L + \Gamma_{31}^M + \Gamma_{21}^R, \nonumber  \\
\dot{\rho}_{22} & = & 0 = \Gamma_{62}^L + \Gamma_{42}^M + \Gamma_{12}^R,\nonumber \\
\dot{\rho}_{33} & = & 0 = \Gamma_{73}^L + \Gamma_{13}^M + \Gamma_{43}^R, \nonumber \\
\dot{\rho}_{44} & = & 0 = \Gamma_{84}^L + \Gamma_{24}^M + \Gamma_{34}^R, \nonumber \\
\dot{\rho}_{55} & = & 0 = \Gamma_{15}^L + \Gamma_{75}^M + \Gamma_{65}^R,  \\
\dot{\rho}_{66} & = & 0 = \Gamma_{26}^L + \Gamma_{86}^M + \Gamma_{56}^R, \nonumber \\
\dot{\rho}_{77} & = & 0 = \Gamma_{37}^L + \Gamma_{57}^M + \Gamma_{87}^R, \nonumber \\
\dot{\rho}_{88} & = & 0 = \Gamma_{48}^L + \Gamma_{68}^M + \Gamma_{78}^R. \nonumber
\end{eqnarray}
These eight equations are not independent since their sum is 0. In order to solve the system for $\rho_{ii}$, one adds the condition $Tr[\rho]=1$ whose resolution provides all state occupation probabilities as well as the currents $J_P$.

We are now going to show that such a system is able to make a thermal transistor analogous to an electronic one. Let us recall that in an electronic bipolar transistor, such as a PNP or a NPN transistor, a current imposed at the base can modulate, switch or amplify the collector and emitter currents. Switching, modulation, and amplification are therefore the caracteristics that must be present in order to have a transistor.  We are going to show here that it is possible to control $J_L$ or $J_R$ by slightly changing $J_M$. The situation is the following : the left and right TLS are both connected to thermal baths, whose respective temperatures $T_L$ and $T_R$ are fixed. The third bath at temperature $T_M$ controls the fluxes $J_L$ and $J_R$ with the help of the current $J_M$ injected into the system. The dynamical amplification factor $\alpha$, defined as
\begin{equation}
\label{ }
\alpha_{L,R}=\frac{\partial J_{L,R}}{\partial J_M}.
\end{equation}
measures the transistor ability to amplify a small heat flux variation at the base (M).
If a small change in $J_M$ induces a large one in $J_L$ or $J_R$, i.e. $|\alpha_{L,R}|>1$, then the thermal transistor effect exists.
The system presented here exhibit many parameters : the frequencies $\omega_P$, $\omega_{PQ}$ and the temperatures $T_L$ and $T_R$.  The last temperature $T_M$, that is taken here between  $T_L$ and $T_R$, controls the transistor properties and is related to the current $J_M$. The number of parameters involved can be reduced by choosing a situation that will not change the physics of the system but will allow a good understanding of the physical phenomena involved. We therefore restrict our analysis to a case for which $\omega_{LM}=\omega_{MR}=\Delta$, whereas $\omega_{RL}$ and the three TLS energies  are equal to 0. As shown below, this configuration provides a good transistor effect, easy to handle with simple calculations. The transistor effect disappears when the three couplings are equal (symmetric configuration), but it still occurs and can even be optimized if the three TLS energies are nonzero \cite{Joulain16}. The operating temperature $T_L$ is taken so that $e^{-\Delta/T_L}\ll1$ ($T_L/\Delta\lesssim 0.25$), whereas $e^{-\Delta/T_R}\ll e^{-\Delta/T_L}$ ($T_R/\Delta\lesssim 0.0625$).

\begin{figure}
\begin{center}
\includegraphics[width=7cm]{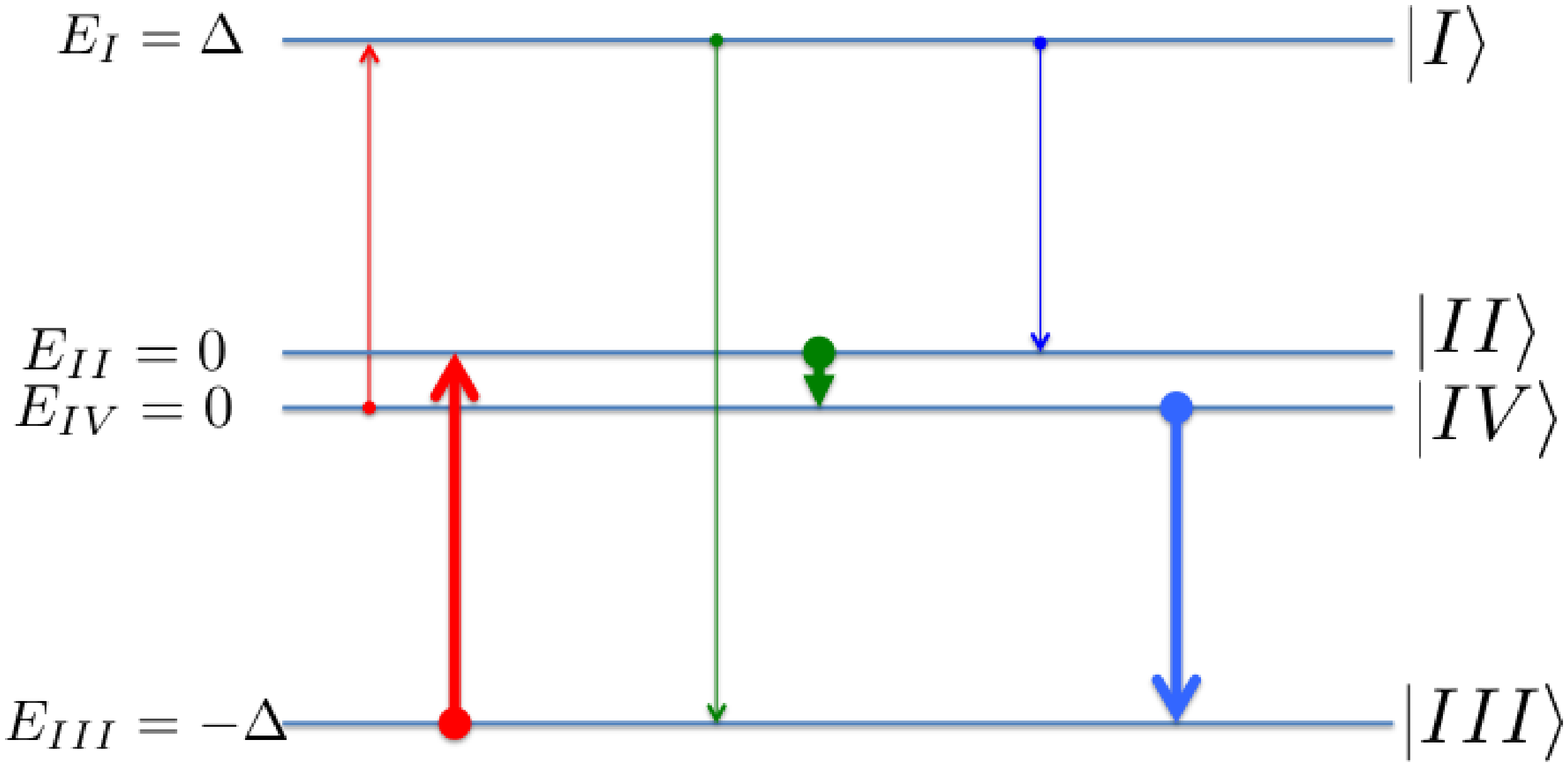}
\caption{Energy levels for $\omega_L=\omega_M=\omega_R=0$, $\omega_{RL}=0$, and $\omega_{LM}=\omega_{MR}=\Delta$. There are four states ($\left|I\right>$, $\left|II\right>$, $\left|III\right>$, and $\left|IV\right>$ but  three energy levels since $E_{II}=E_{IV}=0$. The arrows indicate the net decaying rate between the states due to bath $L$ (red), bath $M$ (green), and bath $R$ (blue)  for $T_L=0.1\Delta$, $T_R=0.01\Delta$, and $T_M=0.05\Delta$.}
\label{Energ_Levels}
\end{center}
\end{figure}

Under these conditions, the system states are degenerated 2 by 2. There are now only 4 states and 3 energy levels (see Fig. \ref{Energ_Levels}).The states $\left|1\right>$ and $\left|8\right>$ are now state $\left|I\right>$, $\left|2\right>$ and $\left|7\right>$ state $\left|II\right>$, $\left|3\right>$ and $\left|6\right>$ state $\left|III\right>$, and $\left|4\right>$ and $\left|5\right>$ state $\left|IV\right>$. One can define the new density matrix elements $\rho_I=\rho_{11}+\rho_{88}$, $\rho_{II}=\rho_{22}+\rho_{77}$, $\rho_{III}=\rho_{33}+\rho_{66}$, and $\rho_{IV}=\rho_{44}+\rho_{55}$. Using the net decaying rates between the states, the three currents read
\begin{eqnarray}
J_L & = & -\Delta\left[\Gamma^L_{I-IV}+\Gamma^L_{II-III}\right] \nonumber\\
J_M & = & -2\Delta\Gamma^M_{I-III}\\
J_R & = &   -\Delta\left[\Gamma^R_{I-II}+\Gamma^R_{IV-III}\right] \nonumber
\end{eqnarray}

\begin{figure}
\begin{center}
\includegraphics[width=8cm]{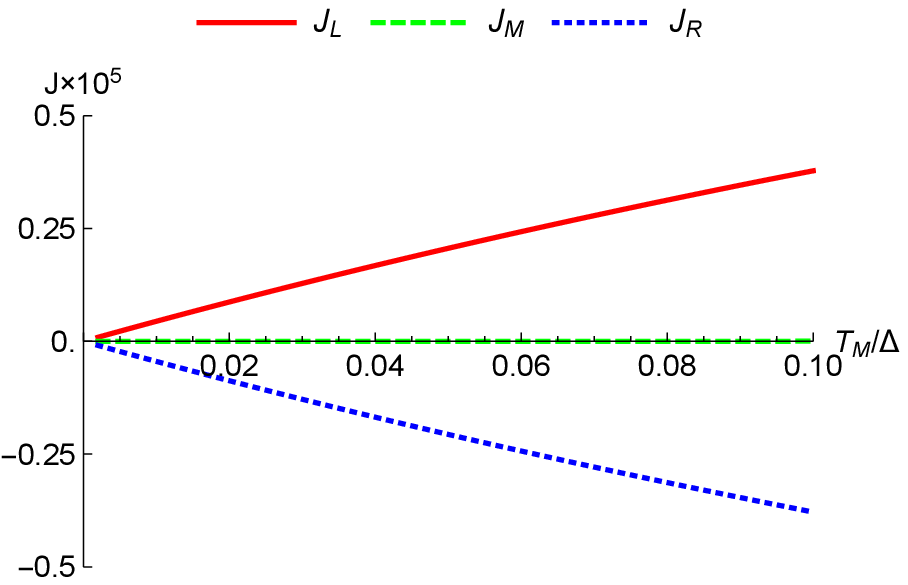}
\includegraphics[width=8cm]{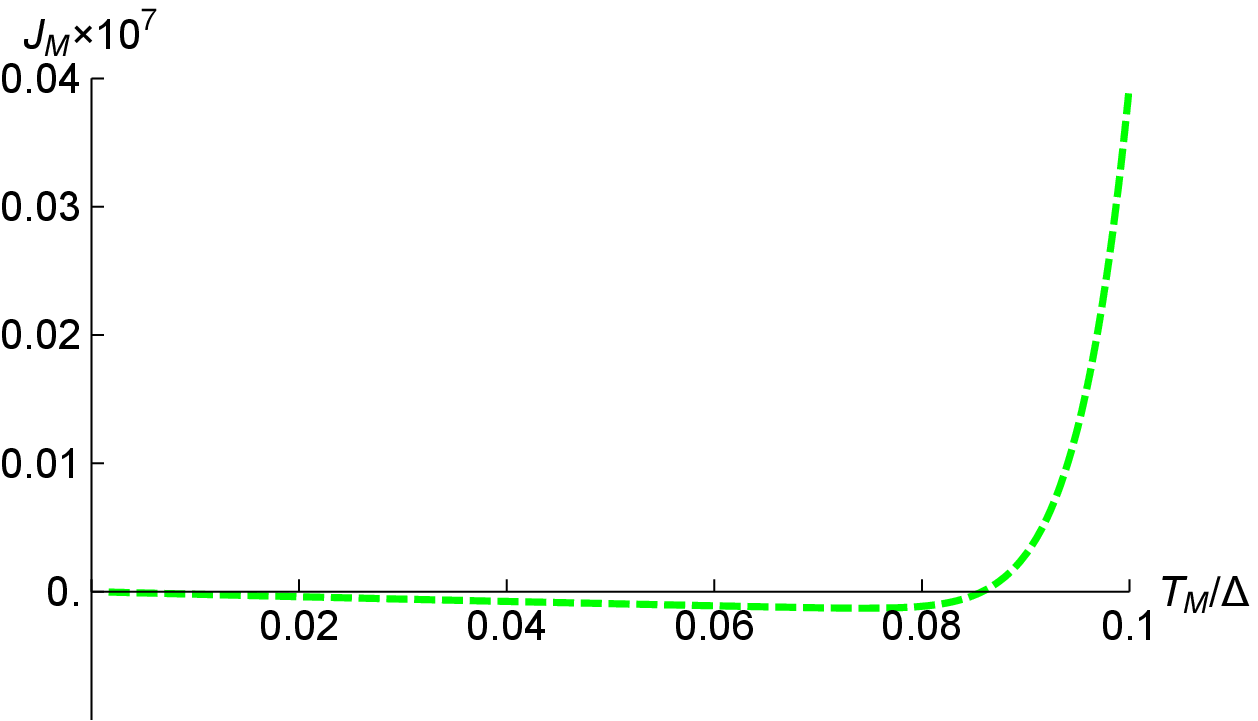}
\caption{Uppern: thermal currents $J_L$, $J_M$, and $J_R$  versus $T_M$ for $\omega_L=\omega_M=\omega_R=0$, $\omega_{RL}=0$, $\omega_{LM}=\omega_{MR}=\Delta$, $T_L=0.1\Delta$, and $T_R=0.01\Delta$. Lower : thermal current $J_M$ versus $T_M$. }
\label{three_currents}
\end{center}
\end{figure}

Transitions between the different states are illustrated in Fig. \ref{Energ_Levels}, for $T_L/\Delta=0.1$, $T_R/\Delta=0.01$, and $T_M/\Delta=0.05$. The arrow directions show the transition direction whereas its width is related to the decay time.  We see that energy exchanges are mainly dominated by the $III-II$ and $IV-III$ transitions. One therefore expects $J_R$ and $J_L$ to be larger than $J_M$. This is illustrated in Fig. \ref{three_currents}, where $J_L$, $J_M$, and $J_R$ are represented versus $T_M$, for $T_L/\Delta=0.1$ and $T_R/\Delta=0.01$. 
$J_L$ and $J_R$ increase linearly with $T_M$, at low temperature, and behave sublinearly as $T_M$ approaches $T_L$. Note that over the whole range, $J_M$ remains lower than $J_L$ and $J_R$, as expected. Thus, $T_M$ will be controlled by changing slightly the current $J_M$: a tiny change of $J_M$ can modify $J_L$ and $J_R$ in a larger proportion. Moreover, $J_L$ and $J_R$ are switched off when $J_M$ approaches 0, for small temperatures $T_M$ : the three TLS system exhibits the transistor switching property. One also remarks that the $J_M$ slope is larger than the ones of $J_L$ and $J_R$ over a large part of the temperature range. Given the definition of the amplification factor $\alpha$, the thermal currents slopes are essential to figure out amplification. 
\begin{figure}
\begin{center}
\includegraphics[width=8cm]{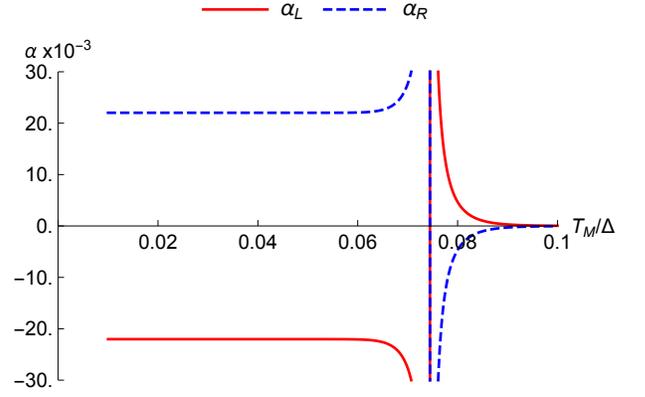}
\caption{Amplification factors $\alpha_L$ (red) and $\alpha_R$ (dashed blue) versus $T_M$ for $\omega_L=\omega_M=\omega_R=0$, $\omega_{RL}=0$, $\omega_{LM}=\omega_{MR}=\Delta$, $T_L=0.1\Delta$ and $T_R=0.01\Delta$.}
\label{alpha}
\end{center}
\end{figure}

In Fig. \ref{alpha}, we plot the two amplification coefficients $\alpha_L$ and $\alpha_R$ versus temperature $T_M$. We see that at low $T_M$, $\alpha$ remains much larger than 1 (around $2.2\times10^4$). One also notes that $\alpha$ diverges for a certain value of the temperature for which $J_M$ has a minimum. This occurs for $T_M\simeq 0.07444\Delta$. In these conditions, an infinitely small change in $J_M$ makes a change in $J_L$ and $J_R$. As $T_M$ approaches $T_L$, the amplification factor drastically decreases to reach values below 1, i.e a regime where we cannot speak anymore of a transistor effect. Note also that, in between,  there exists a temperature for which $J_M=0$. This is the temperature at which the bath $M$ is at thermal equilibrium with the system since it does not inject any thermal current in it. At this temperature ($T_M\simeq 0.08581\Delta$), $J_L=-J_R=3.325\times10^{-6}$. Amplification still occurs since $\alpha_L=831$ and $\alpha_R=-832$.

All these observations can be explained by examining carefully the populations and currents expressions. In the present case, if we limit the calculation to first order of approximations on $e^{-\Delta/T_L}$ and $e^{-\Delta/T_M}$,  one can roughly estimate the populations by
\begin{eqnarray}
\rho_I & \simeq & \frac{e^{-2\Delta/T_M}}{2}+\frac{T_M}{4\Delta+8T_M}e^{-2\Delta/T_L} \label{rho1},\\
\rho_{II} & \simeq & \frac{\Delta+T_M}{\Delta+2T_M}e^{-\Delta/T_L}\label{rho2},\\
\rho_{III} & \simeq & 1-e^{-\Delta/T_L}\label{rho3},\\
\rho_{IV} & \simeq & \frac{T_M}{\Delta+2T_M}e^{-\Delta/T_L}\label{rho4}.
\end{eqnarray}
$\rho_{III}$ remains very close to 1 and $\rho_{II}$ to 10$^{-2}$. $\rho_{I}$ and $\rho_{IV}$
are much lower but change by 1 to 2 orders of magnitude with temperature. 

We now explicitly present the three thermal currents expressions and their dependance with temperature which is the core of our study.
\begin{eqnarray}
J_L & \simeq & -J_R \simeq \frac{\Delta^2 T_M e^{-\Delta/T_L}}{\Delta+2T_M}, \\
J_M & \simeq &\Delta^2\left[-\frac{T_M}{\Delta+2T_M}e^{-2\Delta/T_L}+2 e^{-2\Delta/T_M}\right].
\end{eqnarray}

These formula are in accordance with the linear dependence of the thermal currents for small values of $T_M$. Note also that  $J_L$ and $J_R$ seems to be driven by $\rho_{IV}$, the state population at the intermediate energy ($E_{IV}=0$) when we fully look at their expressions and to (\ref{rho4}). Examining the authorized transitions, one expects $J_M$ to be driven by the population of the most energetic state, i.e., $\rho_I$. The main difference between $\rho_{IV}$ and $\rho_I$ is the temperature dependence, which is linear in one case and exponential ($e^{-2\Delta/T}$) in the other one. The result is that even when $T_M$ is close to $T_L$, $\rho_I$ remains low. Therefore, $J_M$ keeps low values in the whole temperature range due to the low values of $\rho_I$.
If we look more carefully at $J_M$, one notices that it is the sum of two terms. The first one is roughly linear on $T_M$. It is similar to the one that appears in $\rho_{IV}$. $J_M$ depends on the population of state $IV$, which also influences the population of state $I$ with the transition $IV-I$. The increase of $\rho_{IV}$ with $T_M$ makes easier the $IV-I$ transition, and raises $\rho_I$. This increases the decaying of state $I$ through the $I-III$ transition. This term is negative and decreases as $T_M$ increases. This can be seen as a negative differential resistance since a decreasing of $J_M$ (cooling in $M$) corresponds to an increase of the temperature $T_M$. In this temperature range, one can easily show that the amplification factor $|\alpha_L|\approx|\alpha_R|\approx e^{\Delta/T_L}$($e^{10}=22026.5$). 
A second term in $J_M$, is the classical $e^{-\Delta/T_M}$ Boltzmann factor, which makes the population of state $I$ increase with $T_M$. $J_M$ is a tradeoff between these two terms. At low temperature, the linear term is predominant. As $T_M$ increases, the term $e^{-\Delta/T_M}$ takes over. As a consequence, there is a point where the $\rho_I$ increasing reverses the $I-IV$ transition, so that the $I-III$ transition competes with both the $I-IV$ and $I-II$ transitions. $I-III$ is then reversed. With these two terms competing, there is a temperature for which $J_M$ reaches a minimum and a second temperature where $J_M=0$, as already described. 

One can wonder what are the conditions to obtain the best transistor in the conditions studied here. There are two criteria that will quantify a good transistor. One is the amplification factor and the other one is the intensity of the heat currents at the emitter and the collector ($J_L$ and $J_R$).  Note that the amplification factor depends on $e^{\Delta/T_L}$ and that the currents depends on $e^{-\Delta/T_L}$. Let us also recall that we have assumed up to now that $e^{-\Delta/T_L}\ll 1$. Therefore the best choice to have a transistor with a sufficiently collector or emitter current is to take the lowest $\Delta/T_L$ with the condition $e^{-\Delta/T_L}\ll 1$ and the criteria chosen to fullfill this last conditions (here $\Delta/T_L\approx 5$).

One can summarize the conditions needed for the system to undergo a thermal transistor effect. Two baths (here $L$ and $R$) induce transitions between two highly separated states with an intermediate energy level, whereas the third one ($M$) makes only a transition between the two extremes. This will first make $J_M$ much smaller than $J_L$ and $J_R$, and second, it will set a competition between a direct decay of the highest level to the ground level and a decay via the intermediate one. This competition between the two terms makes the thermal dependance of $J_M$ with $T_M$ slow enough to obtain a high amplification.

\section{Conclusion}

We have shown that coupled TLS linked to thermal reservoirs can make systems exhibiting thermal rectification. In the case of 2 TLS, a thermal diode can be made where one of the entry is set at a certain temperature of the order of the system transition. When the other end of the diode is set at a lower temperature, the system is blocked, whereas it is opened when the temperature is higher. This kind of device can isolate a system from cold sources. In the case of a 3 TLS system, we have shown that it is possible to make a thermal transistor. We found a temperature regime where a thermal current variation imposed at the base generates an amplified variation at the emitter and the collector. This regime is typically such that the temperature corresponds to an energy one order of magnitude smaller  than the coupling energy between the TLS. With this kind of thermal transistor one can expect to modulate or amplify thermal fluxes in nanostructures made up of elementary quantum objects.

\begin{acknowledgments}
This work pertains to the French Government Program ''Investissement d'avenir'' (LABEX INTERACTIFS, ANR-11-LABX-0017-01).
\end{acknowledgments}



\end{document}